\documentclass[aps,12pt,showpacs,nofootinbib,preprint]{revtex4}
\usepackage{graphicx}
\usepackage{amsmath,amsfonts}
\usepackage[cp1251]{inputenc}
\usepackage[T2A]{fontenc}
\usepackage[english]{babel}

\newcommand{\pd}{\partial}
\newcommand{\vc}[1]{\mathrm{\mathbf{#1}}}
\newcommand{\rmd}{\mathrm d}
\newcommand{\Elab}{E_{\rm lab}}
\newcommand{\agev}{\ A\text{GeV}}
\newcommand{\enfrz}{\varepsilon_{\rm frz}}
\newcommand{\Tfrz}{T_{\rm frz}}
\newcommand{\enunit}{\ \text{MeV/fm}^3}
\newcommand{\PRC}{Phys. Rev. C }
\newcommand{\NPA}{Nucl. Phys. A }

\def\lsim{\stackrel{\scriptstyle <}{\phantom{}_{\sim}}}

\begin{document}

\selectlanguage{english}
\begin{center}
{\large\bf  Hadron rapidity spectra  within a hybrid model}\\[2mm]
 A. S. Khvorostukhin$^{*,**1}$ and V. D. Toneev$^*$

{\it $^*$ Joint Institute for Nuclear Research,  141980 Dubna, Russia}

{\it $^{**}$ Institute of Applied Physics, Moldova Academy of Science, MD-2028 Kishineu, Moldova}

 \end{center}

{\small A 2-stage hybrid model is proposed that  joins the fast initial state of interaction, described by
the hadron string dynamics (HSD) model, to subsequent evolution of the expanding system at the second stage, treated within ideal hydrodynamics. The developed hybrid model
is assigned to describe heavy-ion collisions in the energy range of the NICA collider under construction in Dubna. Generally, the model is in reasonable agreement with the available data on proton rapidity spectra.
However, reproducing proton rapidity spectra, our hybrid model cannot describe the rapidity distributions of pions. The model should be improved by taking into consideration viscosity effects at the hydrodynamical stage
of system evolution.
}
\\[2mm]
{\small PACS numbers: 25.75.Ag, 25.75.-q\hspace{5mm}{\it Keywords:} heavy-ion collisions, hydrodynamics}

\newpage
\section{Introduction}

Application of hydrodynamics to high-energy nuclear collisions has a long and vivid history which began almost 65 years ago with Landau's original work ~\cite{La53}. In this history a lot of papers has been written covering
a wide spectrum of various and important problems of hydrodynamics. Modern phenomenological status of hydrodynamics is well reflected in recent review-articles   (for example, see~\cite{KH03,HS13,GJS13,JH15,DKK16}).

Hydrodynamics is a collective model of nuclear motion characterized by such physics parameters as temperature, pressure, equation of state (EoS), transport coefficients and so on.  Unfortunately, the direct access to  this information is impossible since the only available experimental information is contained in observable spectra of particles. Hydrodynamics as a theoretical model is just call for relation of observables to thermodynamic properties of excited nuclear matter. As any model, hydrodynamics has its applicability range. The main condition of applicability assumes that the mean free path of a quasiparticle in an excited compressed matter is smaller than the size of this system. It is evident that in nuclear collisions  this condition may be violated in rarefied matter at the initial stage of interaction, in peripheral collisions of nuclei and/or at the final stage of
hydrodynamical expansion.  In this respect, the initial state in the hydrodynamic approach, i.e. space distributions in the energy density, charge density and velocity field, is postulated or calculated within other dynamical models (like Glauber model). The subsequent second interaction stage is described actually by hydrodynamics of dense matter where the key role is played by the equation of state with account for a possible realization of a phase transition of hadrons into a quark phase.

The ultimate aim of this work is the development of a multistage hybrid hydrokinetic approach to heavy-ion collisions in the range of moderate collision energies  $\sqrt s \lsim$ 10 GeV, which are planned to be reached in realizing the projects of heavy-ion collider NICA (Dubna)~ \cite{SST06} and heavy-ion accelerator FAIR (Darmstadt)~\cite{FAIR}. The ideology of our model completely coincides with the hybrid hydro+UrQMD  approach which has been successfully  developed in recent years. The theoretical difficulty lies in the fact that one should construct a model describing available data in a unified way in the whole energy range considered and having a predictive power. A simple use of codes/methods developed for ultrarelativistic energies does not always  lead in any case to satisfactory results. So far the only hydrodynamic model, which well describe a variety of experimental data in this energy range is the three-fluid hydro model~\cite{IRT06}.

\section{Hybrid model}

\subsection{Kinetic stage -- hadron string dynamics model HSD \cite{HSD,PHSD}}

Since the hydrodynamic equations are equations in partial derivatives, one needs to define initial conditions. In hybrid models these conditiions can be obtained from calculations in a kinetic model. This allows one to take into account the nonequilibrium evolution of the system in the initial state of collision. As a model  we choose the hadron string dynamics (HSD) model~\cite{HSD,PHSD}, which describes a large variety of experimental data in the energy range of interest $E_{\rm lab} = 2-50\agev$.

In this paper, as a basis we consider both  the AGS ($E_{\rm lab} = 6$ and 10.7$\agev$), and SPS energy ($E_{\rm lab} =40\agev$, where $E_{\rm lab}$ is the kinetic energy of the bombarding nucleus in the laboratory system). Accordingly, in the AGS  case the calculations are performed for Au+Au collisions, while in the SPS case - for the Pb+Pb system. Therefore, the nucleus radius $R$ slightly differs for different energies.

To get smooth distributions in the density of energy and a baryon number in the initial state, which are averaged
 over a  multiplicity  set of events.  For the initial state the averaging over $5\cdot10^4$ events is used.

The transition from the kinetic description to the hydrodynamic one occurs at some time moment  $t_{\rm start}$. It is assumed that by this time the excited system is close to an equilibrium state that may be characterized by conserving quantites, entropy or the ratio of entropy to the baryon charge~\cite{ST07}. In Ref.~\cite{HPetersen} on the basis of the kinetic calculation results it was proposed to parameterize this moment of transition from the kinetic description to the hydrodynamic one as
\begin{align}
\label{tstart}
t_{\rm start}&=\frac{2R}{\gamma v}=\frac{2R}{\sqrt{\gamma^2-1}}=2R\,
\sqrt{\frac{2m_N}{E_{\rm lab}}}\,,
\end{align}
where the time is counted off nuclear touching moment $t_0$.
This choice corresponds to the moment when nuclei have passed completely through each other. The nucleus radius is calculated as follows $R=r_0\,A^{1/3}$, $r_0=1.124\ \text{fm}$.

 \begin{figure}[thb]
\includegraphics[width=90mm,clip]{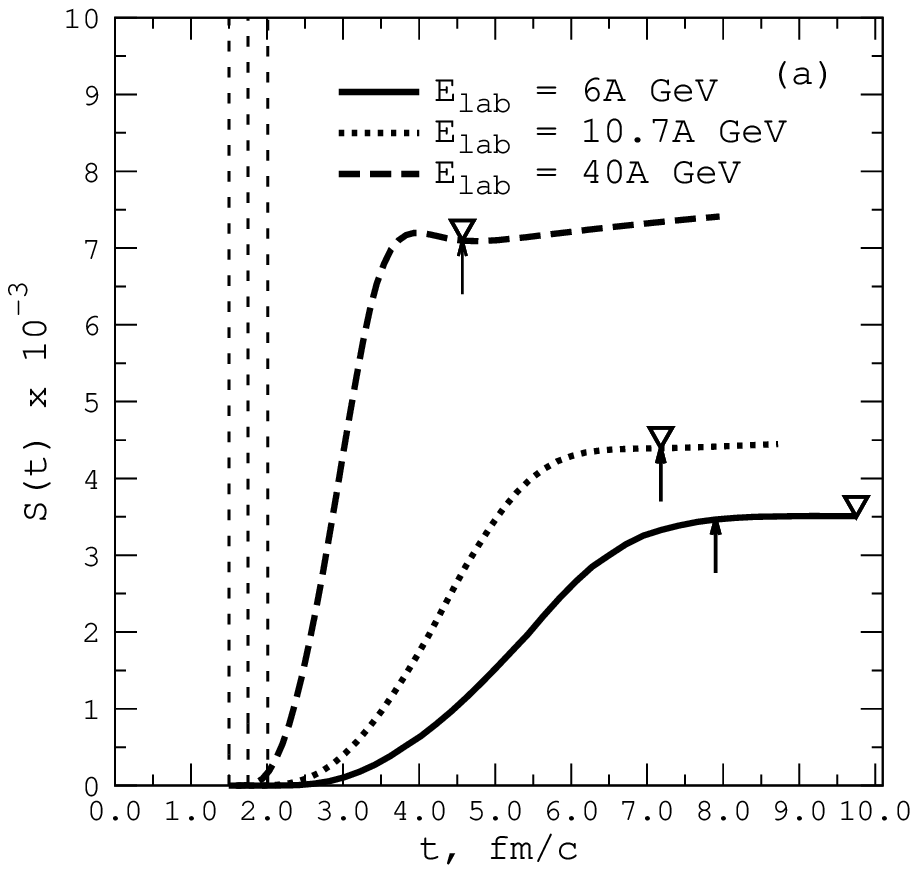}\hspace{-20mm}
\includegraphics[width=90mm,clip]{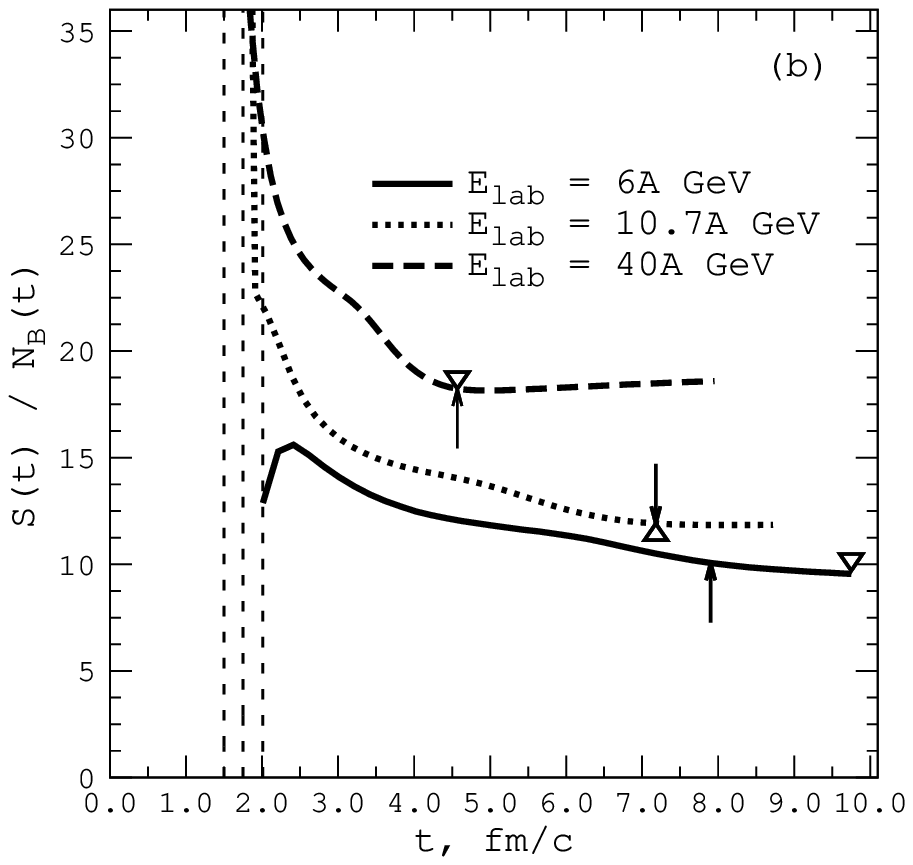}
\caption{Evolution of the total entropy ({\it a}) and the ratio of the total entropy to the baryon number ({\it b}) in central collisions at Pb+Pb ($E_{\rm lab}=40\agev$) and Au+Au ($E_{\rm lab}=6$ and 10.7$\agev$). The vertical lines show the initial moment of the nucleus interaction $t_0$, the arrows are the chosen time for transition to the hydro description
$t_{\rm start}$ (see text), the time  $t_{\rm start}$, calculated according to Eq. (\ref{tstart}), is marked by the triangles.} \label{Stot40}
\end{figure}

Figure~\ref{Stot40} shows how the time estimated according to Eq. (\ref{tstart}) (shown by triangles) correlates with the flattening  moment of the entropy $S(t)$ (panel { \it a}) and the ratio $S(t)/N_B(t)$
(panel {\it b}). Here only particles that have suffered interactions are included; so in the initial collision time  $t_0$, marked by the vertical dashed lines, we have $S=0$ and $N_B=0$. Since the entropy is  mainly generated by pions carring no baryonic charge, the maximum of the function $S(t)/N_B(t)$ is reached soon after the time  $t_0$ and then the function goes down steadily. As is seen, depending on the collision energy, the entropy flattening starts either earlier than the ratio of the entropy to baryon charge (at the AGS energy) or  simultaneously with it (at $\Elab=40\agev$). In that case, Eq.~(\ref{tstart}) appreciably overestimates the time of approaching equilibrium at $6\agev$. This effect is quite understandable if one takes into account that in the transition to lower energy the nucleon-nucleon cross section grows, nuclei are getting less transparent and hence thermalization comes earlier than in a simple geometric estimate of the time moment assuming that nuclei pass freely through each other.

As for the use of a criterium for the beginning of the hydrodynamic interaction stage, we take the constancy of the ratio  $S(t)/N_B(t)$. In Fig.~\ref{Stot40}, the start time of the hydro stage $t_{\rm start}$, estimated as a flattening moment of the ratio $S(t)/N_B(t)$, is shown by the arrows.  For energies 10.7 and 40 $\agev$, this time moment coincides  with  $t_{\rm start}$ calculated according to (\ref{tstart}). Proceeding to lower energies the difference between these two estimates increases.

Certainly, our choice of the transition time to hydrodynamics is more complicated  and less unambiguous than that obtained by the direct calculation of (\ref{tstart}). However, it allows one to easily take into account the effect of nuclear opacity.

As for  the HSD model, the nucleons in colliding nuclei are distributed randomly in the nuclear volume, the touching moment of nuclei is changed from event to event  by some tenths of  fermi. The influence of this effect on observables is nonessential. For the touching moment $t_0$ we take the value for which nuclei are touched  in a rather large number of events.

\subsection{Hydrodynamic stage}

The hydrodynamic equations express simply the conservation laws for the energy-momentum and baryon charge. The ideal hydrodynamics assumes that matter is a local equilibrium state without any dissipative effects. Then the system evolution is described by the following equations~\cite{hydroabout}:
\begin{align}
\label{hydrobase}
    \pd_\mu T^{\mu\nu}&=0,\quad \pd_\mu J^\mu=0,
\end{align}
where the energy-momentum tensor $T^{\mu\nu}$ and the vector of the baryon current $J^\mu$ are
\begin{align}
\label{Tmunu}
    T^{\mu\nu}=(\varepsilon+P)u^\mu u^\nu-Pg^{\mu\nu}, \quad
    J^\mu=n\, u^\mu.
\end{align}
Here $u^\mu = \gamma(1,\vc{v})$ is the vector of the 4-velocity of liquid, $\vc{v}$ is the 3-velocity, the Lorenz-factor is $\gamma=(1-v^2)^{-1/2}$, quantities $\varepsilon$, $n$, $P$ are the energy density, baryon density and pressure in the local reference frame and   $g^{\mu\nu}={\rm diag}(1,-1,-1,-1)$ is the metric tensor.
Eqs. (\ref{hydrobase}) should be completed by EoS $P=P(\varepsilon,n)$, then the system becomes closed.

The local reference frame is a system in which the energy-momentum tensor has the diagonal form. It is possible to show~\cite{hydroabout} that in the ideal hydrodynamics the 4-vector of the entropy current equals
\begin{align}
    s^\mu&= s\,u^\mu,
\end{align}
where $s$ is the entropy density in the local reference frame. The entropy $S$ and the baryon charge $N_B$ of the system are calculated by the integration  of $s^0$ and $J^0$ over the system volume, respectively.

 After substitution of the tensor $T^{\mu\nu}$ and current vector  $J^\mu$ from Eq.(\ref{Tmunu}), Eqs. (\ref{hydrobase}) contain only 5 independent quantities. Their numerical solution together with EoS  $P=P(\varepsilon,n)$ allows one to find the 3 velocity components of the liquid together with the energy density and baryon density in the local reference frame~\cite{SHASTARischke}.
Hydro equations (\ref{hydrobase}) are reduced to a special form and are solved by means of the SHASTA (the SHarp and Smooth Transport Algorithm) algorithm \cite{SHASTA,SHASTARischke}. The SHASTA code is simple in realization and has a rather presice and well-tested algorithm.  A detailed description of the numerical scheme used, where the differential of the generalized pressure is taken in a simplified form, according to \cite{SHASTARischke}, can be found in~\cite{SHASTARischke, Merdeev}. We realized the algorithm in the C/C++ language. The code was tested for the well-known cases of one-dimensional hydro solvable analitically: the Bjorken evolution regime and the expansion of  semi-infinite matter into vacuum  for EoS $P=a\varepsilon$, where $a={\rm const}\leq 1/3$.

In numerical calculations we used the 3-dimensional grid with the cell size $\rmd x=0.2$ fm and the parameter $\lambda=\rmd t/\rmd x=0.4$, which defines the step in time. The EoS for a hadron gas in the mean field proposed in~\cite{SDM} is used and  the $\sigma$-meson is additionally  included in the model data set~\cite{PDG2014}.

\section{Evaluation of observable: particlization procedure}\label{part}

A special task is the calculation of observables: rapidity distributions, transverse momentum spectra and azimuthal flows of particles. The approximation of ``instantaneous freeze-out'' is frequently used in hydrodynamics: it is assumed  that at some space-time hypersurface  there occurs an instantaneous transition from local equilibrium specifying hydrodynamics to collisionless  particle expansion. In these models, the calculation is completed when all cells are frozen. Sometimes it is postulated that the existence of some freezed-out cells do not essentially influence the dynamics of other parts of the system (as an example, see further isothermal and isoenergetic freeze-out). However, there are models where such influence is accounted for some or other method, for example, see~\cite{IRT06}.  A common feature of all these models is the absence of the third (again nonequilibrium) stage of nuclear collision which takes into consideration possible particle rescatterings after fireball expansion when hydrodynamics is not applicable because the particle mean free path becomes too long. In this paper, we construct a two-stage model neglecting posthydrodynamic rescattering. The straightforward method to calculate observables is the application of the Cooper-Frye formulae:

\begin{align}
\label{CooperFrye}
E\frac{\rmd^3N_a}{\rmd p^3}&=\frac{g_a}{(2\pi)^3}\int\rmd\sigma_\nu \frac{p^\nu}{e^{\beta(p^\nu u_\nu-\mu_a)}\pm1},
\end{align}
where $p^\mu=(E,\vc{p})$ is the particle 4-impulse, $\beta=1/T$ -- the inverse local temperature, $g_a$ -- degeneration factor of ``a''\ particle,  $\mu_a$ -- the chemical potential, similar for all particles of the given sort, $\rmd\sigma_\mu=n_\mu\rmd^3\sigma$ -- an element of the space-time freeze-out hypersurface with the normal  $n_\mu$. The plus and minus signs in Eq.~(\ref{CooperFrye}) correspond to the fermion and boson, respectively.
Besides the ``thermal'' contribution calculated by Eq.~(\ref{CooperFrye}), the contributions from the resonance decay are also considered.

To define an integration hypersurface on which the ``instantaneous'' transition from a liquid fluid to freely expanding particles occurs, we use the CORNELIUS algorithm, described in~\cite{Huovinen}, whose realization is available freely. We consider different freeze-out scenarios: 1) \underline {isochronous}, when the calculation is completed  at the given time moment;  2) \underline {isothermal} (iso-$T$), when the cell is frozen if its temperature is less or equals the freeze-out temperature $T\leq\Tfrz$, and 3) \underline
{isoenergetic} (iso-$\varepsilon$), which is  entirely analogous to the isothermal one where the energy density plays the role of temperature. In the last two versions, the frozen cells     are not excluded from the calculation after writing up them into a file and may influence the numerical solution, but this effect is nonessential.

In the Cooper-Frye calculations, only two- and three-body decays in the zero-width approximation are considered. The two-body resonance decays are calculated analytically, see Application B in~\cite{2bodydecays}. The three-body decays are reduced to the two-body ones by substitution of the particle with the mass $M=m_2+m_3$ instead of two particles with the masses $m_2,\ m_3$.

However, below we shall use another approach allowing to speed up appreciably (by an order of magnitude) the numerical calculations. If the hypersurface is known, one may make the inverse transition from the fluid set to quasiparticles using the Monte-Carlo method -- so called particlization. According to our algorithm, the number of particles in each cell is calculated following~\cite{HPetersen}, while the 4-momentum is sampled according to~\cite{FastMC}. Since the Cartesian coordinate system is used, the particle space coordinates are random quantities homogeneously distributed within the given cell and do not affect the momentum distribution. The contribution of space-like cells is ignored. After generation, the two- and three-body resonance decays are taken into account in the zero-width approximation. For the SPS energy only strong and electromagnetic decays but for the AGS energy all decays (besides the decay of charged and long-lived kaons) are considered.

To test our generator, the simulated distributions are compared to direct calculations of spectra, according to the Cooper-Frye formulae~\cite{CooperFrye}.

\begin{figure}[thb]
\includegraphics[width=90mm,clip]{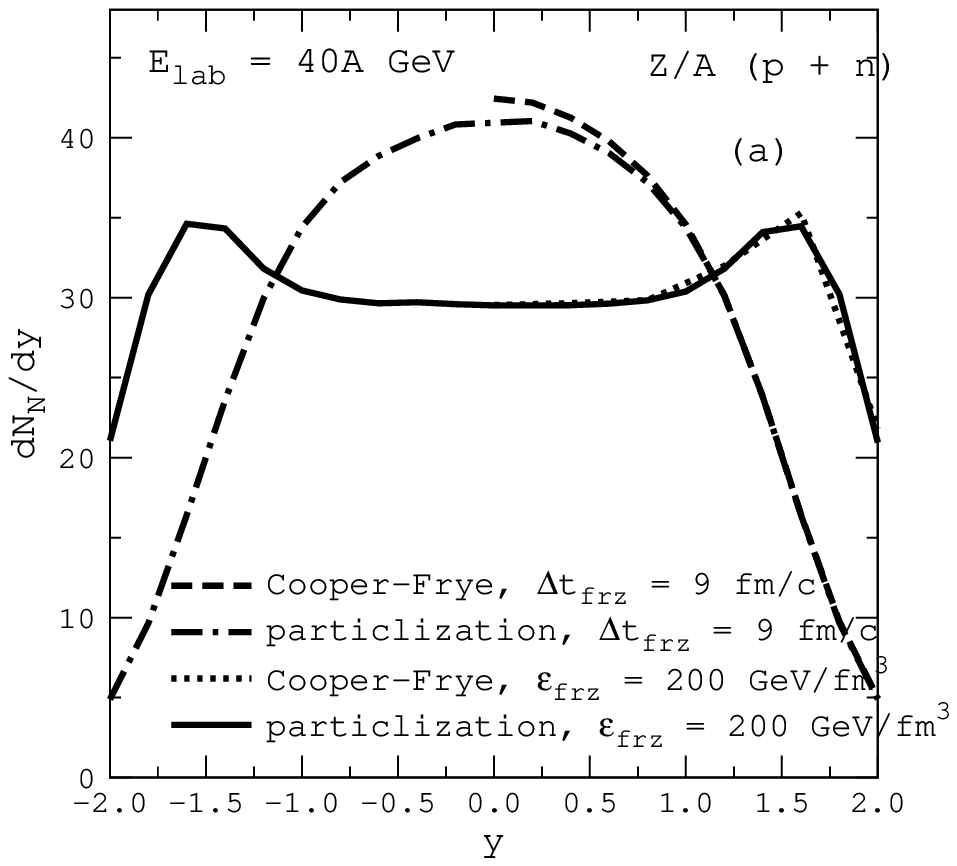}
\hspace{-20mm}
\includegraphics[width=90mm,clip]{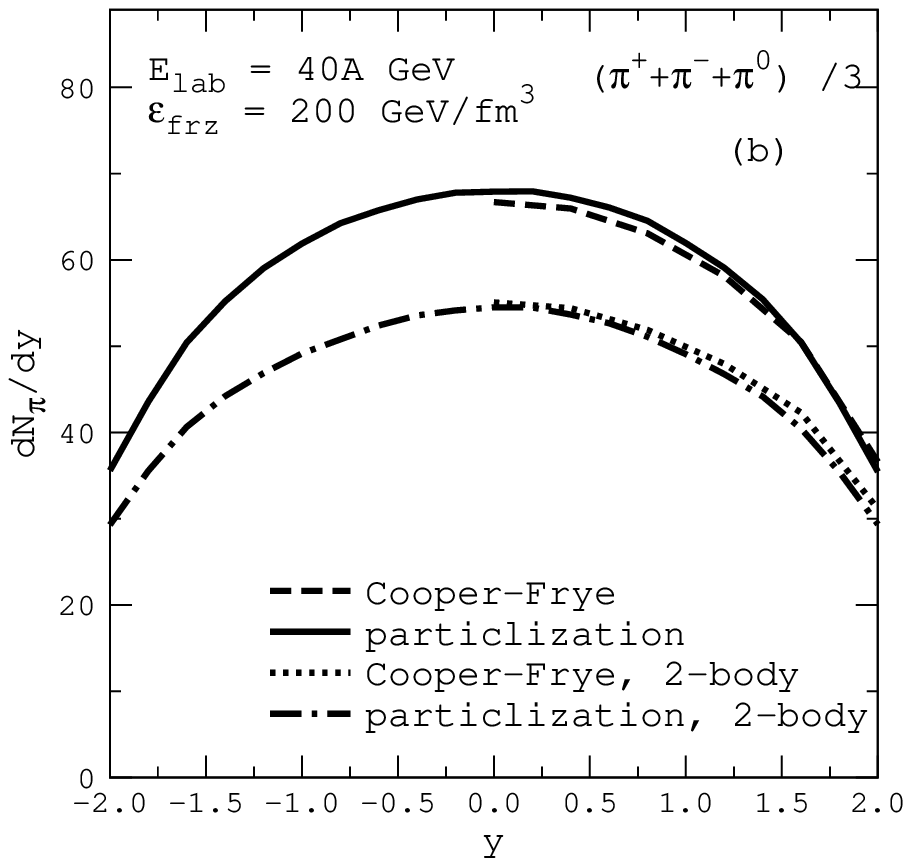}
\caption{ Comparison of rapidity spectra calculated by the Cooper-Frye  to Monte-Carlo
results of particlization at  $\Elab=40\agev$ and $b=2.5$ fm for ({\it a}) nucleons and  ({\it b}) pions. Panel ({\it a}) ~ the solid and dashed lines are particlization and direct calculations, according to Eq.~(\ref{CooperFrye}), respectively, evaluated for isoenergetic freeze-out at $\enfrz=200\enunit$; the dash-dotted lines are particlization and Cooper-Frye for the initial state from \cite{Merdeev} and isochronous freeze-our at $\Delta t_{\rm frz}=9$ fm. Panel ({\it b})~the solid and dashed lines are particlization and Cooper-Frye
for the iso-$\varepsilon$ scenario with $\enfrz = 200\enunit$, the
dash-dotted and dotted lines are the same distributions but only two-body decays are included.} \label{MCtest}
\end{figure}

In Fig.~\ref{MCtest}{\it a}, the comparison of the results is presented for nucleons with the the isochronous, when $\rmd\sigma_\mu=\delta_{\mu,0}\rmd^3x$, and iso-$\varepsilon$
freeze-out scenarios. A similar comparison for pions in the iso-$\varepsilon$ scenario is given in Fig.~\ref{MCtest}{\it b}. It is seen that the Cooper-Frye method and our algorithm provide practically coinciding curves. In the case with isochronous freeze-out, the initial state from Ref.~ \cite{Merdeev} is constructed as a sum of density distributions of two cold nuclei coming toward each other. In Fig.~\ref{MCtest}{\it b}, a special case is presented for pions when only two-body decays are included which, substantially underestimates the pion yield. The agreement of the directly calculated spectra with the  particlization results allows us to apply this procedure for calculation of observables instead of application of
Eq.~(\ref{CooperFrye}).

Since our hydrodynamic equations do not include a separate equation for conserving an electric charge current, we use the symmetric EoS. Therefore, there is a question how to estimate correctly the particle fraction with the given electric charge among particles with the same mass
(for example, the ratio of protons to the total number of nucleons).
As seen from Fig.~\ref{MCtest}$a$, the isochronous scenario differs noticeably from two others: multiplying a number of nucleons by the isotopic factor $Z/A$ we get $dN/dy$ at $y=0$
for protons essentially larger (closer to experiment) than for iso$\varepsilon$ freeze-out.
Therefore, an additional isotopic factor depends also on the freeze-out scenario. Thus, below we will compare  the averaged experimental  proton multiplicity with the calculated
$(n_p+n_n)/2$ for the iso-$\varepsilon$ and iso-$T$ scenarios.

 As to the pion yield,  in all cases we take the isospin average $(n_{\pi^+}+n_{\pi^-}+n_{\pi^0})/3$.  Our calculations show that such averaging is closer to the yield of $n_{\pi^+}$. One should note that charge asymmetry is observed clearly in nuclear experiments in the NICA energy range. For example, for pions created at the SIS accelerator $n_{\pi{-}}/n_{{\pi+}}\sim 1.7$. This ratio slowly decreases with the energy growth reaching the unit at the energy about 150$\agev$~\cite{pi-ratio}.

At energies of interest the ratio  $n_{\pi^-}/n_{\pi^+}$ exceeds remarkable 1. However, any
version of the hadronic transport model HSD overestimates pion multiplicity in the NICA energy range (see for example~\cite{MCPB16}). This overestimate of pions results in a too low kaon-to-pion ratio $K^+/\pi^+$. This problem is actively discussed in the last years and frequently associated with the signal of a possible quark-hadron phase transition.

\section{Confronting results with different freeze-out scenarios}

It is evident that the assumption on isochronous freeze-out is not realistic and can be used only for test calculations or as an intermediate stage. In realistic models where observable quantities are calculated on a "frozen"\ hypersurface, a constant temperature of the energy density  scenarios is used. To construct a model having predictive power, one needs to choose a scenario and a method for calculating its parameters depending on collision energy.

In our analysis of central nucleus-nucleus collisions, on default, we use the impact parameter $b = 1$ fm at all energies under discussion since, as a rule,
 it gives results close to experiments.
\begin{figure}[thb]
\includegraphics[width=90mm,clip]{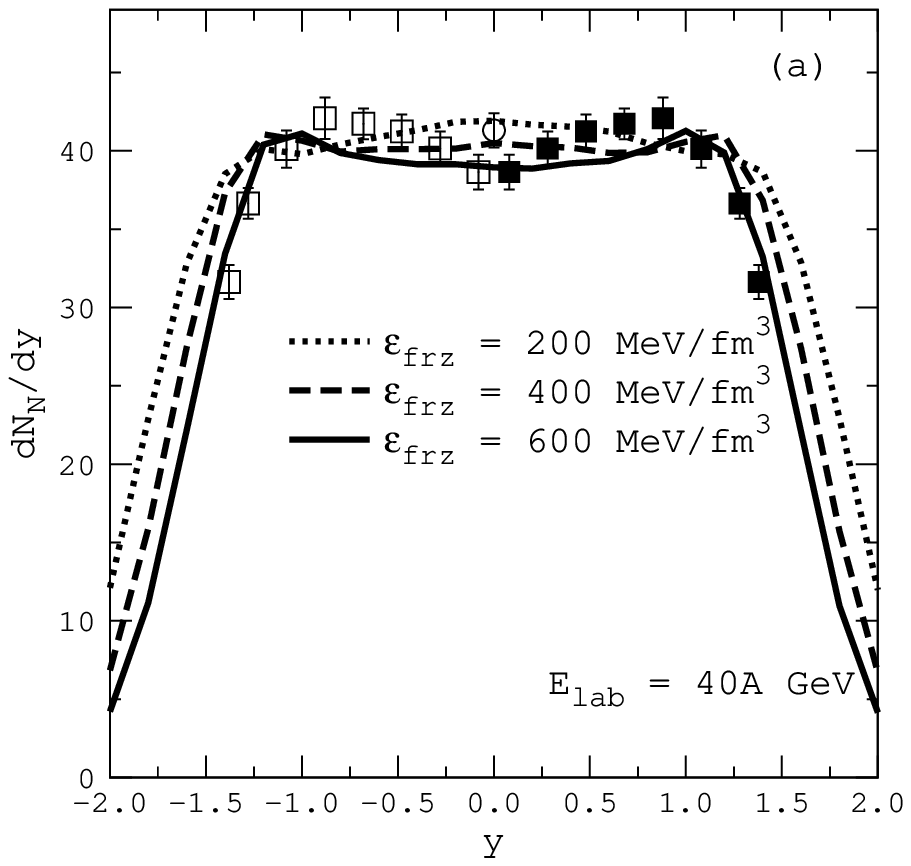}
\hspace{-2cm}
\includegraphics[width=90mm,clip]{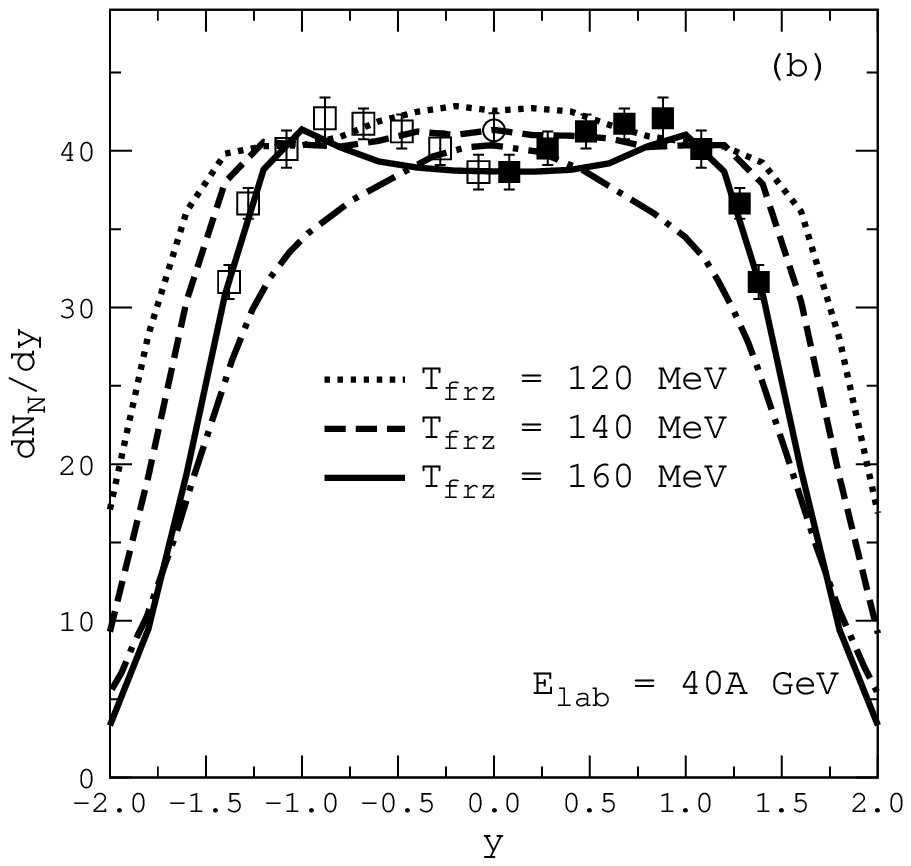}
\includegraphics[width=90mm,clip]{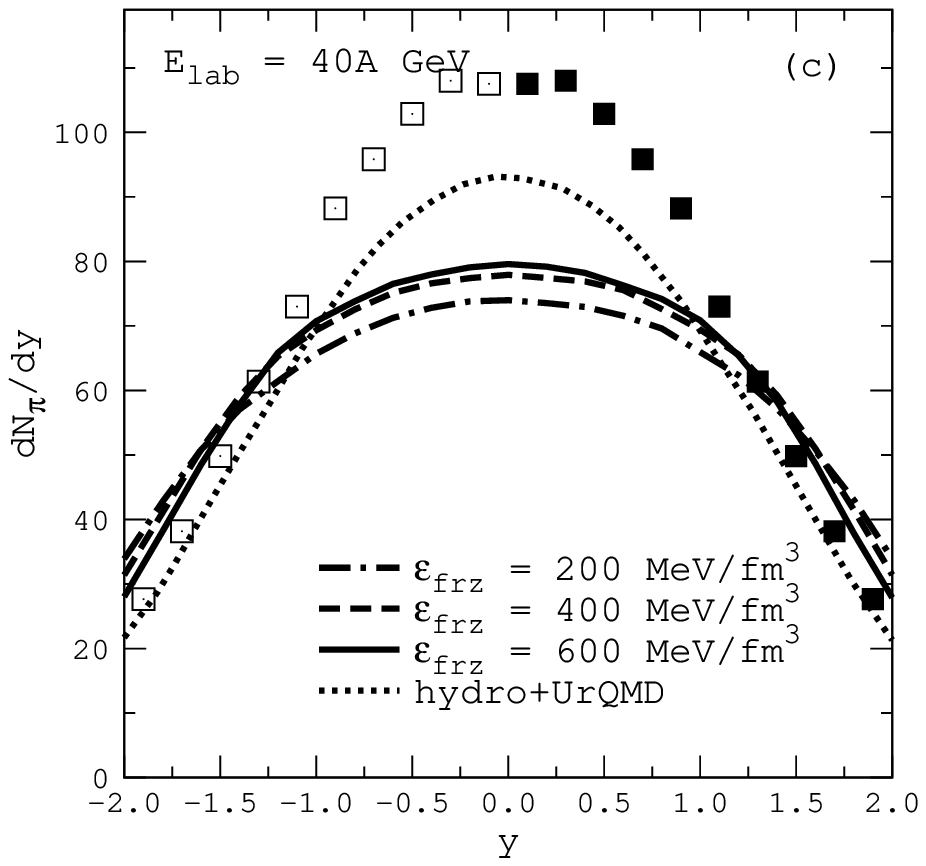}
\hspace{-2cm}
\includegraphics[width=90mm,clip]{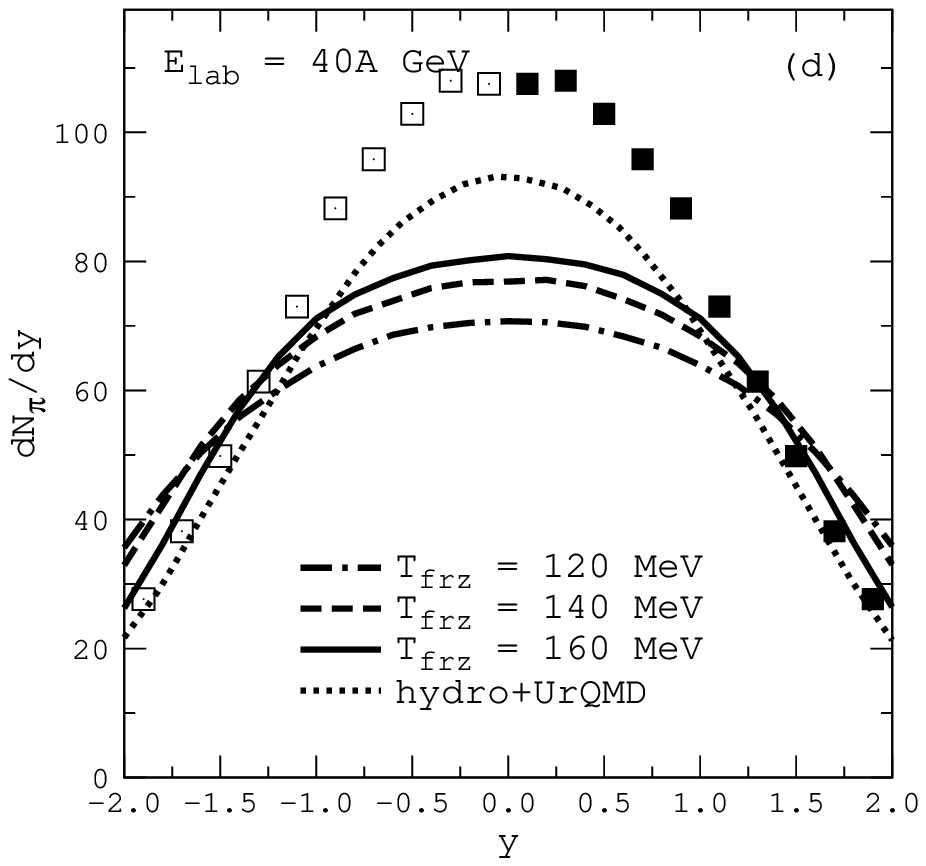}
\caption{Rapidity distributions for protons (top) and pions (bottom) in the central ($b=$ 1fm)
Pb+Pb collisions at  $\Elab=40\agev$  in isoenergetic (left) and isothermal (right)
freeze-out. The dash-dotted line is the result for isochronous freeze-out from Ref.~\cite{Merdeev}; the dotted line -- hybrid hydro+UrQMD result~\cite{KHPB}.
Experimental points for protons are taken from~\cite{SPSN1}
(squares) and \cite{SPSN2} (circles), for pions -- from~\cite{SPSpiK}). In all figures the open symbols are obtained by reflection of measured points with respect to the line $y=0$.}
\label{pdndy_E40}
\end{figure}

\begin{figure}[thb]
\includegraphics[width=90mm,clip]{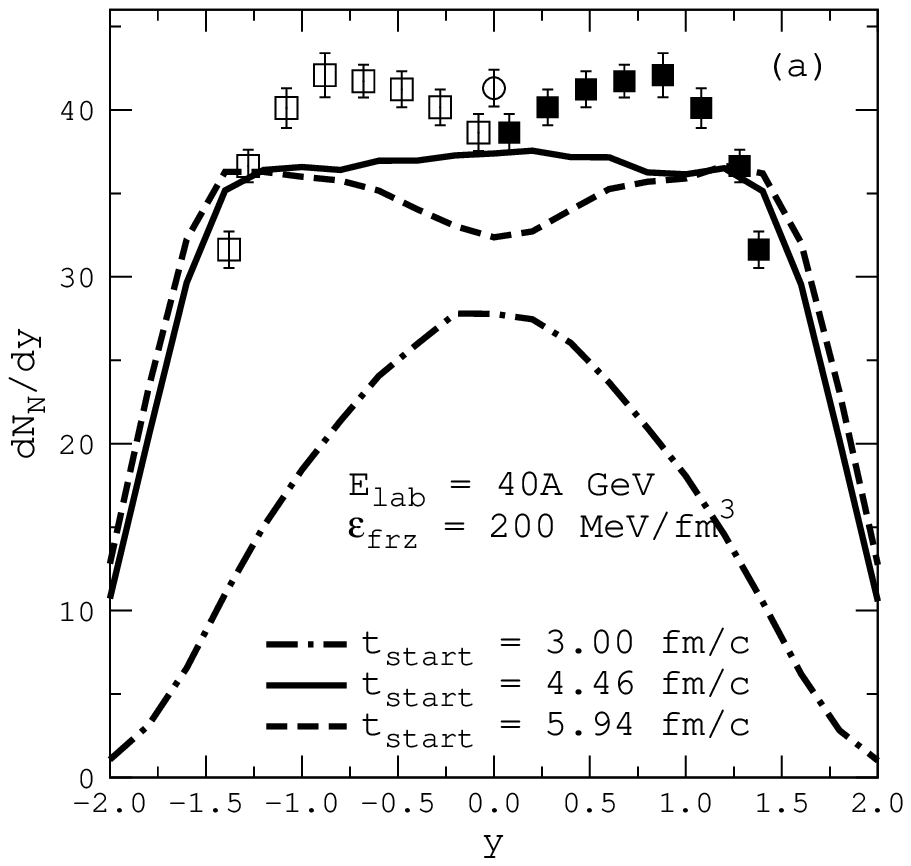}
\hspace{-2cm}
\includegraphics[width=90mm,clip]{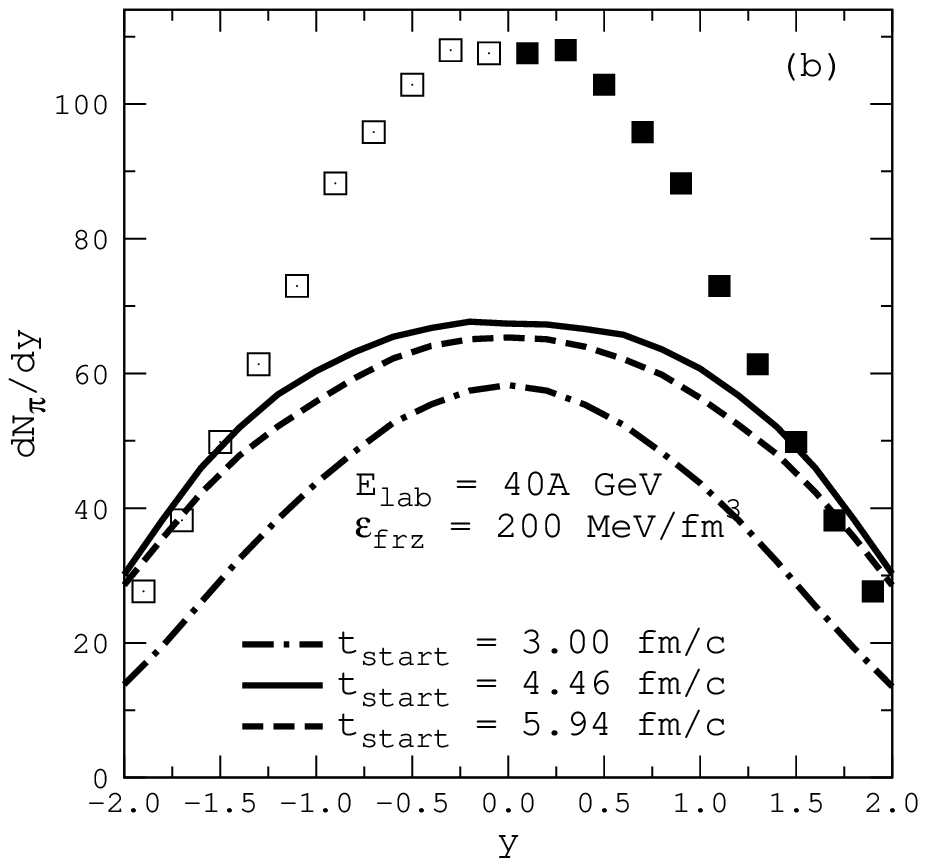}
\caption{Rapidity distributions ({\it a}) protons and ({\it b}) pions at various
 $t_{\rm start}$ for Pb+Pb collisions at $\Elab=40\agev$ and $b=2.5$ fm with isoenergetic freeze-out $\enfrz=200\enunit$. Experimental points are the same as in Fig.~\ref{pdndy_E40}.}
\label{ptime_E40}
\end{figure}

It is shown at the top of Fig.~\ref{pdndy_E40} that the rapidity proton spectra depend on({\it a}) the freeze-out energy density  $\varepsilon_{\rm frz}$ and ({\it b})the freeze-out temperature $T_{\rm frz}$ at the bombarding energy $\Elab = 40\agev$. It is seen that to reproduce the two-hump structure, one needs to take rather large values of the parameters  $\varepsilon_{\rm frz}$ or  $T_{\rm frz}$. It is of interest that in contrast with the kinetic  HSD model, the hybrid model  has no problem with the description of proton spectra at large rapidity.

For comparison, in the same figure we give the results~\cite{Merdeev} obtained in the model with isochronous freeze-out. It is of interest to note that these results demonstrate the point that to get a two-hump structure, the  two-phase EoS is necessary while our results show that the choice of the initial state and method/parameter of freeze-out plays not less important role.  Such ambiguity between the choice of a proper initial state or EoS was also noted in~\cite{Huov1997}. Since the model~\cite{Merdeev} with the two-phase EoS predicts the two-hump structure at $\Elab=10.7\agev$ as well, one may conclude that this method for constructing the initial state is too simplified.

It is shown in Fig.~\ref{ptime_E40} how the rapidity distributions of protons and pions depend on the hydro stage initial time $t_{start}$ in the case of $E_{lab}=40\agev$ and $b=$ 2.5 fm   the isoenergetic freeze-out. It is seen that in out hybrid model the two-hump structure arises also at later time transition to hydrodynamics. This confirms the conclusion on an important role of the initial state in transition to hydrodynamics.

The pion rapidity distributions for  $\Elab=40\agev$ presented in Fig.~\ref{pdndy_E40} and Fig.~\ref{ptime_E40} show that the distribution height in the rapidity range
$y \approx 0$ depends on both the freeze-out parameter and  $t_{\rm start}$; moreover Eq.~(\ref{tstart}) describing flattening the $S/N_B$ gives the maximum in this distribution. However, we did not succeed in reproducing experimental pion spectra in either the isothermal or   isoenergetic scenarios. For comparison, the result of the hybrid version of the UrQMD model  and hydrodynamics without viscosity~\cite{KHPB} is shown (the dotted curve in Fig.~\ref{pdndy_E40}). Our model differs from it only by the point that the initial state here is taken from the UrQMD model and hydrodynamics additionally includes the electric charge conservation  and hence the results should be  quite similar. It is seen that the hydro-UrQMD  model results in lower than experimental data pion rapidity distributions though the maximal value lies closer to the experiment, which is explained by both the use of a different EoS and mainly by the account for electric charge conservation,  which should provide a higher yield of  negative pions measured at the SPS.

\begin{figure}[thb]
\includegraphics[width=62mm,clip]{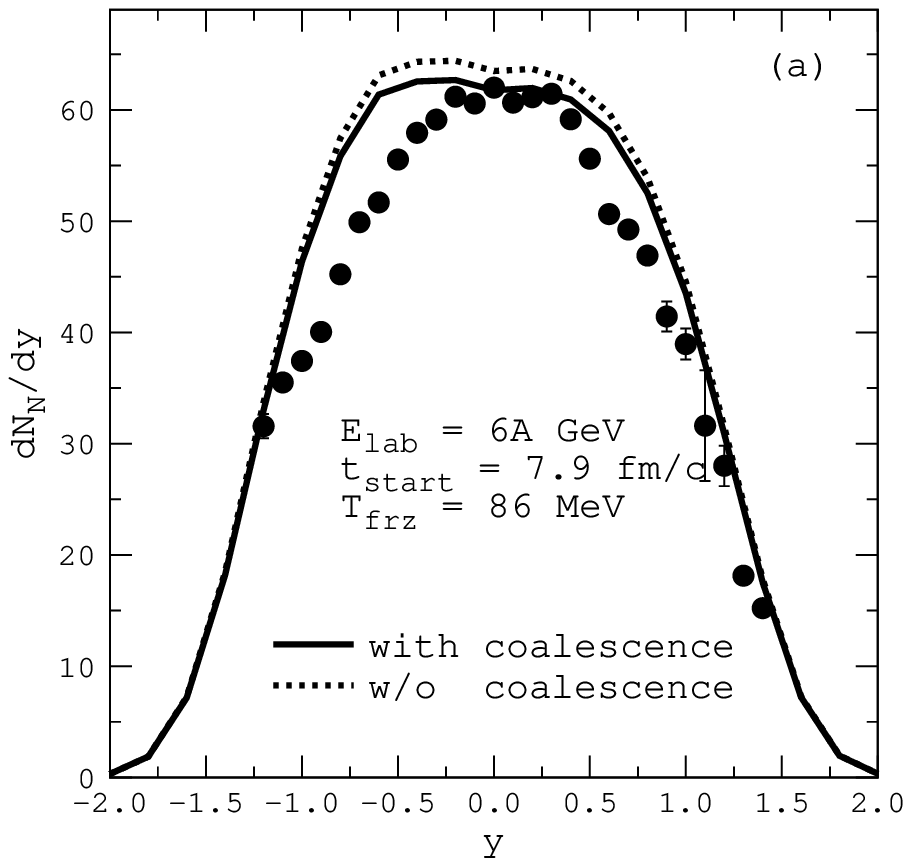}
\hspace{-1.5cm}
\includegraphics[width=62mm,clip]{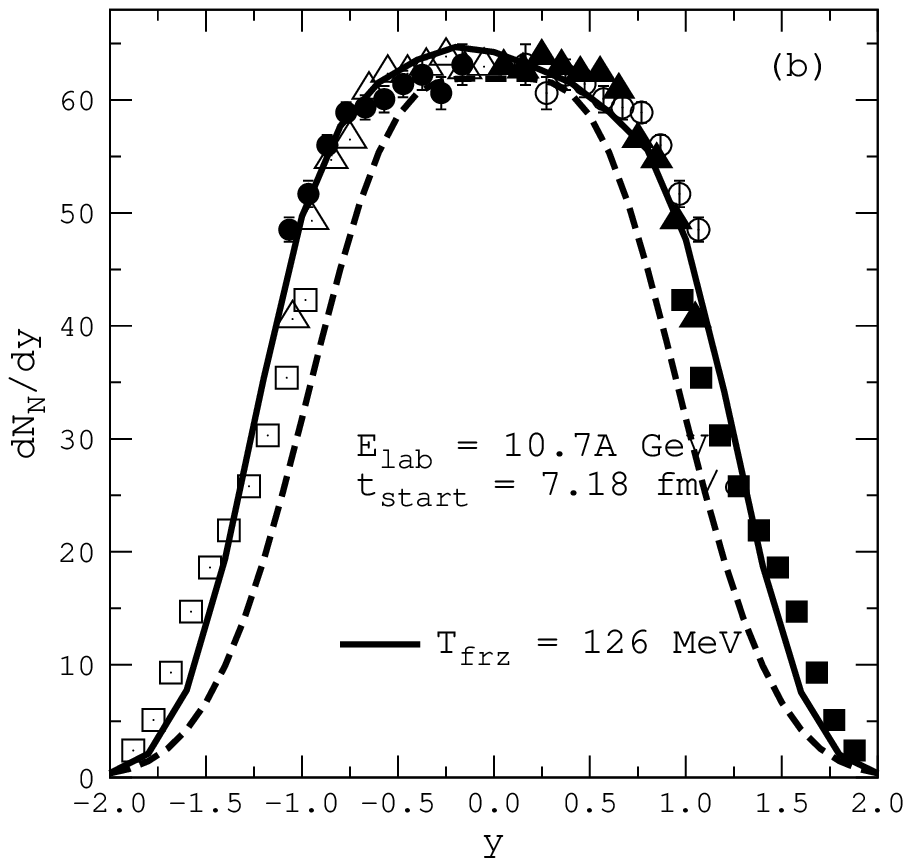}
\hspace{-1.5cm}
\includegraphics[width=62mm,clip]{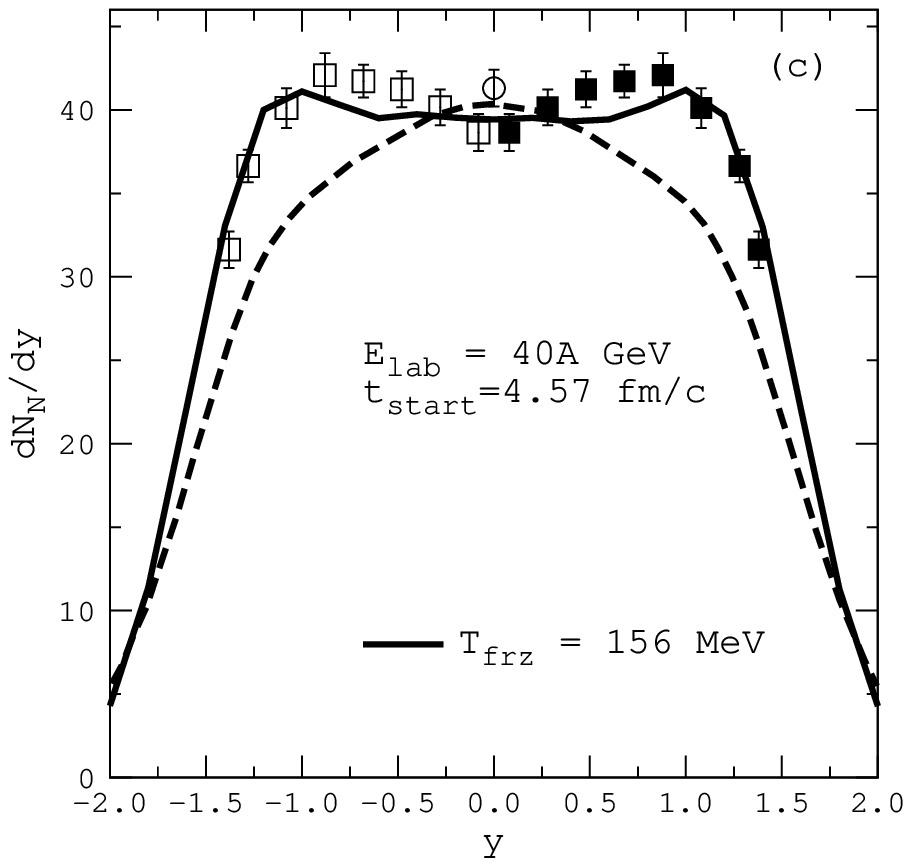}
\caption{ Ultimate proton spectra  within the isothermal scenario  at energies ({\it a}) $\Elab=6\agev$, ({\it b}) $\Elab=10.7\agev$ and ({\it c}) $\Elab=40\agev$. The solid lines are our results, the dashed are the results with isochronous freeze-out from Ref.~\cite{Merdeev} ({\it b, c}). Experimental points for $\Elab=6\agev$ are taken from \cite{E895protons}, for $\Elab=10.7\agev$ --  \cite{PRC57} (triangles), \cite{PRC62}
(squares) and \cite{PRL86} (circles).} \label{Nfinal}
\end{figure}

As has been  noted for the first time in Ref.~\cite{Hi04} and clearly demonstrated in Ref.~\cite{KHPB}, the lack of pions is due to the absence in our model of dissipative effects  which increase the entropy. In addition, one should remember that the SPS data are given for  $\pi^-$ pions, while the isospin averaged calculation results are closer to the number of less abundant $\pi^+$ pions.

Similar consideration within iso-$T$ and iso-$\varepsilon$ scenarios at $\Elab=10.7\agev$ does not result in worse agreement. The protons are close to experimental ones if one chooses  $\Tfrz\sim120$ MeV or $\enfrz\sim200\enunit$. In contrast to the energy $\Elab=40\agev$, in these cases the pion distributions are only slightly below experimental ones.

The confrontation of two energies leads to the conclusion that in both scenarios the freeze-out parameter giving the best description  with experiment depends on the collision energy.
\begin{figure}[thb]
\includegraphics[width=62mm,clip]{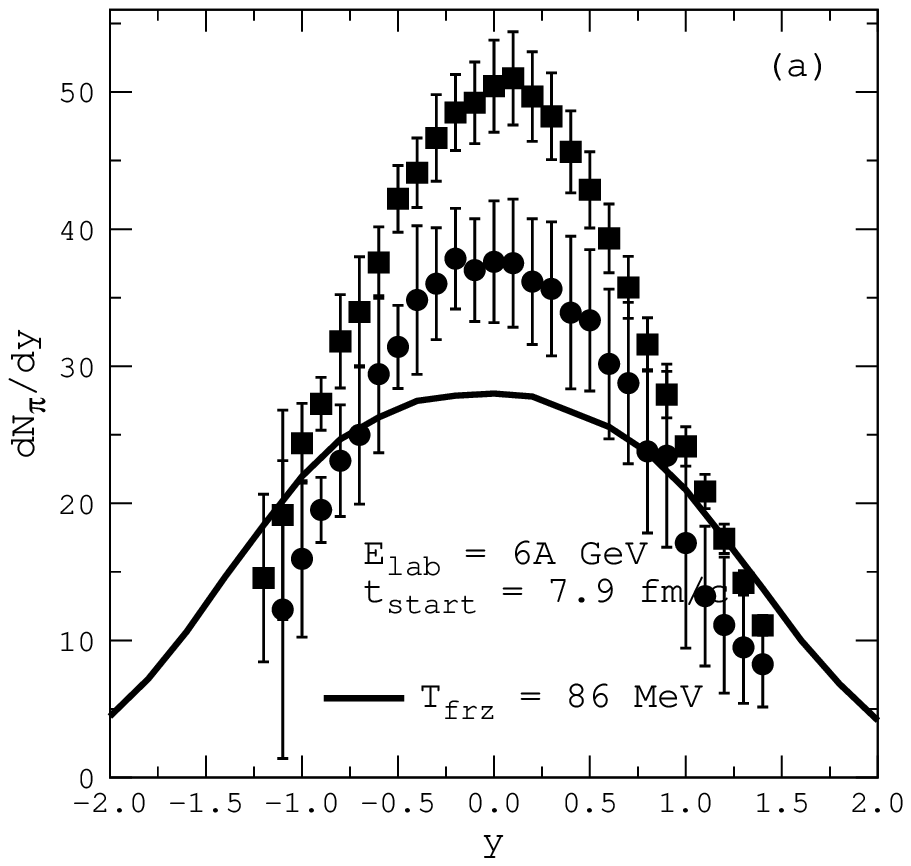}
\hspace{-1.5cm}
\includegraphics[width=62mm,clip]{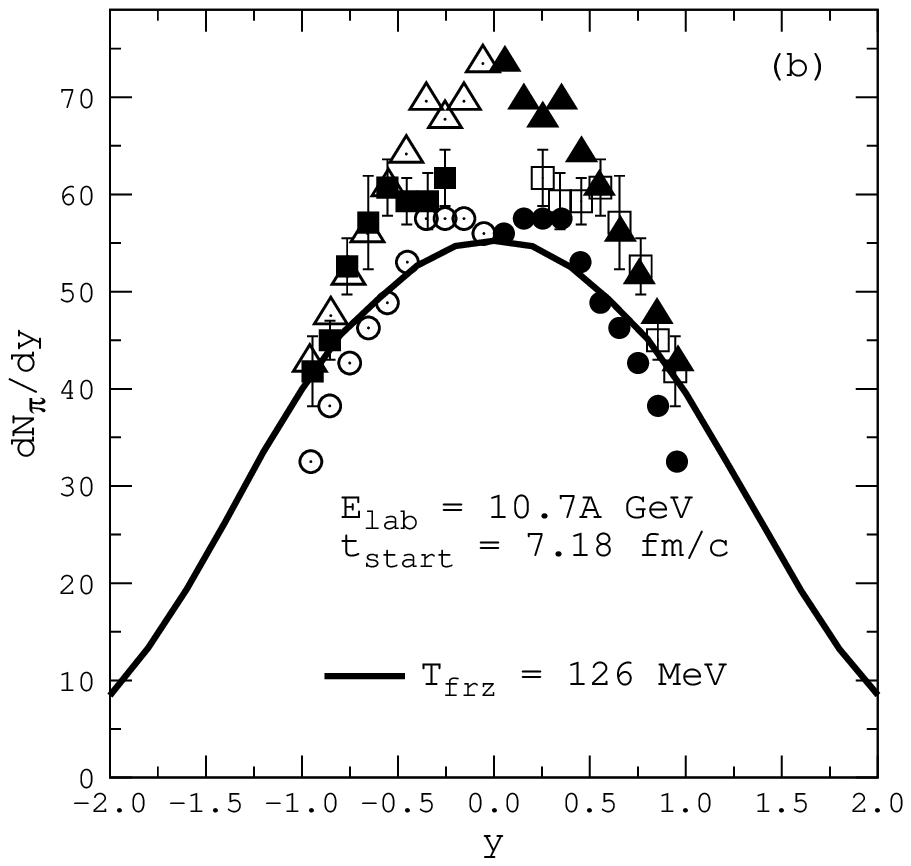}
\hspace{-1.5cm}
\includegraphics[width=62mm,clip]{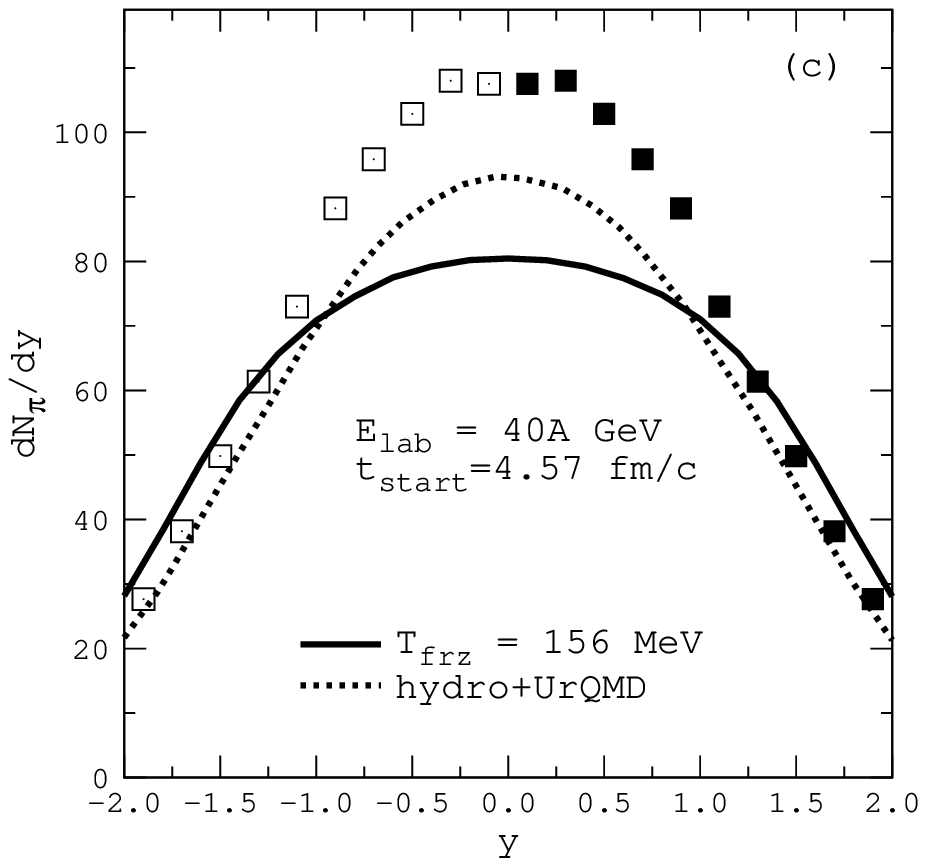}
\caption{Rapidity distribution as in Fig.~\ref{Nfinal} but for pions. The dotted curve corresponds to the hybrid model~\cite{KHPB}. Experimental points for $\Elab=6\agev$ are taken from~\cite{PRC68}, for $\Elab=10.7\agev$ -- from~\cite{PRC59} ($\pi^+$
-- squares)  and \cite{NPA610} ($\pi^-$ -- triangles, $\pi^+$ -- circles).}
\label{pifinal}
\end{figure}
As is known, the phenomenological statistical model of hadron and resonance production~\cite{Andronic} predicts the dependence of the freeze-out temperature $T_{frz}$ on the collision energy.
Thus, the solution is suggested to be to used as input data temperatures obtained from the data analysis within the statistical model. For all energies considered $\Elab=6, 10.7$ and $40\agev$  good agreement with experiment for protons is achieved if to use respectively the values of $T_{frz}=86^{+13}_{-3}$, 124${\pm 7}$ and 156${\pm 11}$~MeV \cite{Andronic} (see. Fig.~\ref{Nfinal}).

Appropriate pion distributions for the same energies are presented in Fig.~\ref{pifinal}.
Our model does not allow us to reproduce them in a regular basis. The calculation results turn out to be close to the experiment only at $\Elab=10.7\agev$, but they are also somewhat  underestimated. The lack of some pions is observed  in the hydro-UrQMD~model~\cite{KHPB}, as follows from Fig.~\ref{pifinal}{\it c}.

Thus, our hybrid model suffers some difficulties in a simultaneous description of the AGS and SPS energy range. The model with isochronous freeze-out  considered above~\cite{Merdeev} also has some problems in attempt to consider these both energies  simultaneously because
one did not succeed in describing
 the shape of proton spectra within a single EoS.

\section{Conclusions}

For describing heavy-ion collisions in the energy range reachable at heavy-ion collider NICA which is under construction in Dubna, the 2-stage hybrid model is proposed which joins together the fast interaction stage considered within the kinetic model of hadron-string dynamics HSD and the subsequent system expansion stage evaluated in terms of ideal hydrodynamics. For this model three versions of the freeze-out scenarios are realized: isochronous, isothermal and isoenergetic. The description of sensitivity to different elements of the hybrid model is illustrated.

In the large, the model is in qualitative satisfactory agreement with experiments on hadron spectral distributions. The 2-stage version allows one to describe the proton spectra  reasonably and even quantitatively within the isothermal scenario if the freeze-out temperatures $\Tfrz$ are taken from the data obtained by a statistical model  processing the measured hadron yield.
It is shown that within the hybrid model the parameters of the two-hump structure in the proton spectra  may be obtained by either increasing the freeze-out temperature or energy density  parameters or by transition to the hydrodynamical stage at a later time. The ideal hydrodynamics with the initial state calculated on the basis of the kinetic HSD model is not able to describe pion rapidity spectra. The reason of this discrepancy is in neglecting the hadron  matter viscosity at the hydrodynamic stage of the system evolution.

It is of interest to mention that the account for viscosity is not the only way to improve the description of pions in hydrodynmics. A similar result may be reached within a multifluid approach, in particular in the 3-fluid hydrodynamics~\cite{IRT06}.  One should note that both the account for viscosity and the 3-fluid approach mean the departure from  the local-equilibrium concept, i.e. from the ideal 1-fluid hydrodynamics. The account for mutual friction in the 3-fluid model brings to the dissipation of the relative motion energy of colliding nuclei and entropy generation, which is then realized through the pion emission at the decay of the third baryonless fluid. The model describes well multiplicity of identified hadrons in the energy range discussed~\cite{IRT06,Iv13}. A number of model parameters is comparable or even less than in hybrid models, and these parameters are clearly fixed. This fact allows one to proceed to a detailed analysis, in particular, to investigate irregularities in proton rapidity spectra and relate them to anomalies in the EoS stipulated by a possible hadron-quark phase transition~\cite{Iv13}.

At moderate energies, important characteristic is the stopping power which is described by the energy fraction transforming into created particles and thereby defining the initial state of the system.  In its turn, the selected energy specifies the nature   (hadronic or quark-gluon) for a subsequent evolution of excited matter. In contrast to a full kinetic approach in HSD and the account of this effect in the 3-fluid model, in hybrid models this fraction is estimated not completely, only at the first fast interaction stage and is governed by the transition time to the hydro description which is the parameter at every collision energy.  Then the local equilibrium is assumed to be established instantly. Such a rough approach  to baryon stopping  is manifested in a maximal way just at the moderate collision energies, since  the nuclear stopping power grows strongly with the energy in the range $\sqrt s \lsim$ 10 GeV and flattens at higher energies~\cite{IRT06}.

  \vspace{3mm}
  \begin{center} {\bf Acknowledgements:}
We much appreciate to D. Rischke and B. Betz for providing us the SHASTA code and E. Bratkovskaya for the HSD code.
We are thankful to  V.~Voronyuk, Yu.~Ivanov, Iu.~Karpenko, A.~Merdeev, L.~Satarov and  G.~Sandukovskaya for useful discussions and constructive remarks.
\end{center}


\begin{thebibliography}{99}

\bibitem{La53}L. Landau, 
Izv. Akad. Nauk Ser.Fiz. {\bf 17}, 5164 (1953).
%
\bibitem{KH03}P. F. Kolb and U. W. Heinz, 
Published in "Quark gluon plasma", 634 
 (2003), Hwa, R.C. (ed.) {\it et al.} [arXiv: nucl-th/0305084].
%
\bibitem{HS13}U. W. Heinz and R. Snellings, 
Annu. Rev. Nucl. Part. Sci. {\bf 63}, 123  (2013) [arXiv: 1301.2826].
%
\bibitem{GJS13} C. Gale, S. Jeon, and B. Schenke, 
Int. J. Mod. Phys. A {\bf28}, 1340011 (2013).
%
\bibitem{JH15}S. Jeon and U. Heinz, 
arXiv: 1503.03931 (2015).
%
\bibitem{DKK16}R. Derradi de Souza, T. Koide and T. Kodama, 
Prog. Part. Nucl. Phys. {\bf 86},  35 (2016) [arXiv: 1506.03863].

%
\bibitem{SST06}A. N. Sissakian, A. S. Sorin and V. D. Toneev, Cong. Proc. C {\bf 060726}, 421 (2006) [arXiv: nucl-th/0608032].
%
\bibitem{FAIR}B. Friman, C. Hohne, J. Knoll, S. Leupold, J. Randrup, R. Rapp, P. Senger, Lect. Notes Phys. {\bf 814}, 1 (2011).
%
\bibitem{IRT06} Yu. B. Ivanov, V. N. Russkikh, and V. D. Toneev, 
    \PRC {\bf73}, 044904 (2006) [arXiv: nucl-th/0503088].
%
\bibitem{HSD} W. Ehehalt and W. Cassing,  
Nucl. Phys. A {\bf602}, 449 (1996);
J. Geiss, W. Cassing, and C. Greiner,  
Nucl. Phys. A {\bf 644}, 107 (1998);
W. Cassing and E. L. Bratkovskaya, 
Phys. Rept. {\bf 308}, 65  (1999).

\bibitem{PHSD} W. Cassing and E. L. Bratkovskaya, 
Nucl. Phys. A {\bf831}, 215 (2009);
Phys. Rev. C {\bf78}, 034919 (2008).
%
\bibitem{ST07} V. V.~Skokov and V. D.~Toneev, 
Yad. Phys. {\bf 70}, 114  (2007) [Phys. Atom. Nucl. {\bf 70}, 109  (2007)].
%
\bibitem{HPetersen} H. Petersen, J. Steinheimer, G. Burau, M. Bleicher, H. St\"{o}cker, 
\PRC {\bf78}, 044901 (2008).
%
\bibitem{hydroabout} L. D. Landau, E. M. Lifschitz, Fluid Mechanics, vol. 6, Addison–Westley, 1959.
%
\bibitem{SHASTARischke} D. H. Rischke, S. Bernard, and J. A.Maruhn, 
    Nucl. Phys. A {\bf595}, 346 (1995).
%
\bibitem{SHASTA} J. P. Boris and D. L. Book, 
J. Comp. Phys. A {\bf11}, 38 (1973);
D. L. Book, J. P. Boris, and K. Hain, 
J. Comp. Phys. A {\bf18}, 248 (1975).
%
\bibitem{Merdeev} A. V. Merdeev, Hydrodynamic modelling of the quark-hadron phase transition, PhD thesisis, Moscow, 2011 (in russian); L. M. Satarov, private communication.
%
\bibitem{SDM} L. M. Satarov, M. N. Dmitriev, and I. N. Mishustin, 
    Phys. Atom. Nucl. {\bf72}, 1390 (2009).

\bibitem{PDG2014} K. A. Olive {\it et al.} (Particle Data Group), Chin. Phys. C {\bf38}, 090001 (2014).

\bibitem{Huovinen} P. Huovinen and H. Petersen, 
Eur. Phys. J. A {\bf 48}, 171 (2012).
%
\bibitem{2bodydecays} L. M. Satarov,  I. N. Mishustin, A. V. Merdeev, 
Phys. Atom. Nucl. {\bf70}, 1773 (2007) [arXiv: hep-ph/0611099].
%
\bibitem{FastMC} N. S. Amelin,  R. Lednicky, T. A. Pocheptsov, I. P. Lokhtin, L. V. Malinina, A. M. Snigirev, Iu. A. Karpenko, Yu. M. Sinyukov, 
\PRC {\bf 74}, 064901 (2006).
%
\bibitem{CooperFrye} F. Cooper and G. Frye, 
    Phys. Rev. D {\bf 10}, 186 (1974).
%
\bibitem{pi-ratio} G. S. F. Stephans, 
in {\it Proceedings of the RHIC/AGS Annual Users' Meeting 2007, Brookhaven, 18-22, June, 2007},
www.bnl.gov/rhic\_ags/users\_meeting/Past\_Meetings/2007/Agenda/Fri/
Stephans\_RHIC\_Users\_2007.pdf .
%
\bibitem{MCPB16} W. Cassing, A. Palmese, P. Moreau, E. L. Bratkovskaya,  
arXiv: 1510.04120 (2015).
%
\bibitem{KHPB} Iu. A. Karpenko, P. Huovinen, H. Petersen, M. Bleicher, 
   \PRC {\bf91}, 064901 (2015) [arXiv: 1502.01978].
%
\bibitem{SPSN1} T. Anticic {\it et al.}, 
    \PRC {\bf83}, 014901 (2011).
%
\bibitem{SPSN2} T. Anticic {\it et al.}, 
    \PRC {\bf69}, 024902 (2004).
%
\bibitem{SPSpiK} S. V. Afanasiev {\it et al.}, 
\PRC {\bf66}, 054902 (2002).
%
\bibitem{Huov1997} J. Sollfrank, P. Huovinen, M. Kataja, P. V. Ruuskanen, M. Prakash, R. Venugopalan , 
\PRC {\bf 55}, 392 (1997) [arXiv: nucl-th/9607029].
%
\bibitem{PRC57} L. Ahle {\it et al.}, 
\PRC {\bf57}, R466 (1988).
%
\bibitem{PRC62} J. Barrette {\it et al.}, 
\PRC {\bf62}, 024901 (2000).
%
\bibitem{PRL86} B. B. Back {\it et al.}, 
Phys. Rev. Lett. {\bf86}, 1970 (2001).
%
\bibitem{PRC68} J. L. Klay {\it et al.}, 
\PRC {\bf 68}, 054905 (2003).
%
\bibitem{PRC59} L. Ahle {\it et al.}, 
    \PRC {\bf59}, 2173 (1999).
%
\bibitem{NPA610} L. Ahle {\it et al.}, 
\NPA {\bf610}, 139c (1996).
%
\bibitem{E895protons} J. L. Klay {\it et al.}, 
Phys. Rev. Lett. {\bf 88}, 102301 (2002).
%
\bibitem{Hi04}  T. Hirano, J. Phys. G: Nucl. Part. Phys. {\bf 30},  S845 (2004).
%
\bibitem{Andronic} A. Andronic, P. Braun-Munzinger, and J. Stachel, 
\NPA {\bf772}, 167 (2006).
%
\bibitem{Iv13}  Yu. B. Ivanov, 
\PRC {\bf87}, 064904 (2013) [arXiv:1302.5766].

\end{thebibliography}
\end{document}